\UseRawInputEncoding 
\documentclass[twocolumn,aps]{revtex4}
\usepackage{latexsym}
\usepackage{amsmath}
\pdfoutput=1
\usepackage{times}
\usepackage{amssymb}
\usepackage{fancyheadings}
\usepackage[T1]{fontenc}
\usepackage{url}
\usepackage{hyperref}
\usepackage{color}
\usepackage{graphicx}
 \usepackage{epsfig}
\usepackage{mathptmx}
\usepackage{bm}
\usepackage{array}
\usepackage{algorithm}
\usepackage{algorithmic}
\usepackage{url}
\usepackage{hyperref}
\usepackage{algorithm}
\usepackage{algorithmic}
\usepackage{color}

\usepackage{lscape}

\begin{document}

\title{Hidden geometry and dynamics of complex networks: Spin
  reversal  in nanoassemblies with pairwise and triangle-based interactions  
}

\author{Bosiljka Tadi\'c$^{1,2}$,  Neelima Gupte$^{3}$}
\vspace*{3mm}
\affiliation{$^1$  Department of Theoretical Physics, Jožef Stefan Institute, Ljubljana, Slovenia}
 \affiliation{$^2$Complexity Science Hub Vienna, Josephstädter Strasse 39,
  Vienna, Austria}
\affiliation{$^3$ Indian Institute of Technology Madras, Chennai, India}

\date{}

\begin{abstract}
\noindent Recent studies of networks representing complex systems from
the brain to social graphs have revealed their higher-order
architecture, which can be described by aggregates of simplexes
(triangles, tetrahedrons, and higher cliques).  Current research aims
at quantifying these hidden geometries by the algebraic topology
methods and deep graph theory and understanding the dynamic processes on simplicial complexes.  Here, we use the recently introduced model
for geometrical self-assembly of cliques to grow nano-networks of
triangles and study the filed-driven spin reversal processes on
them. With the antiferromagnetic interactions between the Ising spins
attached to the nodes, this assembly ideally supports the geometric
frustration, which is recognized as the origin of some new phenomena
in condensed matter physics. In the dynamical model,  a gradual
switching from the pairwise to  triangle-based interactions is
controlled by a parameter. Thus, the spin frustration effects on each
triangle give way to the mesoscopic ordering conditioned by a complex
arrangement of triangles. We show how the balance between these
interactions changes the shape of the hysteresis loop. Meanwhile, the
fluctuations in the accompanying Barkhausen noise exhibit robust
indicators of self-organized criticality, which is induced by the network geometry alone without any magnetic disorder.
\end{abstract}

\maketitle

\section{Introduction\label{sec:intro}}
 The influence of network structure on dynamics in many complex systems
has been demonstrated in numerous studies with detailed analysis of
empirical data and theoretical approaches 
\cite{we-SciRep2018,SD_RevModPhys2008,we_topReview2010,Geometry_KrioukovPRE2010,we_nanonetworks2013,we_PRE2015,Spectra-topologyPRE2017,Spectra_wePRE2019,millan2019synchronization,Neelima_TSchaos2020}. Hence, the origin of structure of many complex networks can be sought in their
evolutionary optimisations towards efficient functioning.
Recently,  numerous studies revealed that complex systems obey higher-order connectivity that can be described by simplexes of different sizes, see a recent overview in \cite{SM-book2017}. Beyond standard graph-theory methods \cite{bb-book}, the aggregate simplicial complexes are suitably described by Q-analysis \cite{Q-analysis-book1982} based on the algebraic topology of graphs \cite{jj-book}.
Some striking examples include human connectomes
\cite{HOC_Brain_weSciRep2019,HOC_Brain_weHubsSciRep2020} and
(multi)-brain networks originating from experimental signals
\cite{we-Brain-PLOS2016,we-Frontiers2018,Neelima_fNMRI2017}. The higher-order
interactions based on simplexes appear as constitutive characteristics
of complex systems from biology \cite{HOC_baudot2019}, materials science \cite{SA_allscales_Science2002,SA_colloids_scales2016,FunctColloids_Nat2012,ProgMatt_review,AT_Materials_Jap2016}
and physics \cite{Geometries_QuantumPRE2015,we_Qnets2016} to the large-scale social  dynamics 
\cite{we_PhysA2015,we_Tags2016,HB-knowNets-BT2019,HOC_socialModelMSmall2020}. 

Current research of complex networks with hidden structures highlights three major directions. Specifically, 
\begin{itemize}
\item exploring empirical data to discover the  higher-order
  structures and the related dynamical patterns in them (Q-analysis,
  persistent homology, topological information data analysis) \cite{SM-book2017,HOC_brainScafold2014,HOC_baudot2019};

\item theoretical investigations of different 
  stochastic processes taking part on such structures to reveal the
  impact of higher-order structures and new phenomena that they may
  induce \cite{HOC_Synchro_ArenasPRL2019,HOC_BHN_weEntropy2020,HOC_socialModelMSmall2020};

\item modelling the assembly of networks with simplicial complexes to control the appearance of specific structural characteristics given altered function or dynamic processes that they can support. For example, pre-formatted groups, described as cliques (full graphs),  are attached to a growing structure during  the geometrically constrained process of self-assembly
\cite{bianconi-SciRep2017,we-SciRep2018,we-PRE2020}. See 
the illustration in Fig.\ \ref{fig:nets}.
\end{itemize}

In materials science, the impact of the complex architecture of nano-materials to their functionality has been a subject of experimental investigations, which are often driven by the specific
requirements for applications. A prominent example is the case of
antiferromagnetic spintronics  \cite{Frustr_AFMspintronicsNatNano2016}. In this context, recent focus is on materials with complex
architecture, like tetraborides \cite{Frustr_tetraborides_SREP2018}
and artificial frustrated systems \cite{Frustr_design2016}.
In such systems, the geometric constraints combined with the sign of
the interactions prevent simultaneous minimisation of the global
energy and each spin pair, leading to the spin frustration
 \cite{Frustr_book2005}. 
The geometric frustration effects have been shown to enable an unusual
magnetic order and new dynamical phenomena in condensed matter physics 
\cite{Frustr_book2015Julich,Frustr_nanostructures2017,Frustr_PRE18,Frustr_AFnanostruct2019}
as well as in the Ising model applications \cite{HOC_RezaBalance2019,YHolovatch_IsingspinAppl2020}. 
 For example, one of the prominent features of spin
frustration is the appearance of  the fractional magnetisation plateaus in the hysteresis loop 
\cite{Frustr_SSlatticeFractMplateus_PRL2012,Frustr_SSlattice_FractMplateus_PLA2014,Frustr_tetraborides_SREP2018}.  
In \cite{HOC_BHN_weEntropy2020},  the nanonetworks of self-assembled
mono-disperse cliques were studied. It was shown that, when endowed
with antiferromagnetic bonds between pairs of spins on triangle faces,
these assemblies provide ideal conditions for geometric
frustration. The size of building simplexes directly affects the shape
of the hysteresis loop, and the fractional magnetisation plateaus.

In this work, we study the spin-reversal dynamics on self-assembled
nanonetworks with simplicial complexes in the presence of
simplex-based interactions. More precisely, the network is
grown by geometrical self-assembly of triangles as pre-formatted
groups of nanoparticles, based on the self-assembly model introduced
in \cite{we-SciRep2018}, and the Ising spin variable is attached to
each node  (nanoparticle). With antiferromagnetic
interactions among pairs of neighbouring spins,  the spin frustration effects are fully supported by the geometry of these simplexes
\cite{HOC_BHN_weEntropy2020}. Here, in addition to the pairwise
interactions, we introduce a simplex-based interaction involving three spins on a triangle. Our motivation is to assess the impact of the higher-order interactions, which are naturally enabled by the structure of the assembly, by balancing them
(via a parameter $\alpha$) against the strong spin frustration effects
promoted by the antiferromagnetic pairwise interactions on the same
structure. We analyse their combined influence on the shape of the
hysteresis loop and the multi-scale structure of the magnetisation
fluctuations during the reversal process.

\section{The self-assembly of simplexes\label{sec:results}} The above mentioned Q-analysis reveals the geometric
constituents of a simplicial complex in a given network. The generative
models, on the other hand, offer a bottom-up approach to grow a
structure with  simplicial complexes controlled by pertinent parameters.
In the prototype model introduced in
\cite{we-SciRep2018}, which we use here, the self-assembly of cliques
is constrained by the geometric compatibility and chemical
affinity factors (see also  \cite{we-PRE2020} for an extended model with defect
cliques). More precisely, a simplex (clique) of the size $n\in[2,10]$
is added by sharing one of its faces, i.e., sub-cliques of the
size $n_q<n$, with an already existing clique in the growing
structure. The faces of the order $q=0,1,2,3 \cdots q_{max}-1$ indicate a single
node, link, triangle, tetrahedron, and so on up to the largest sub-clique, where
$q_{max}=n-1$ is the order of the added clique. The probability of a
clique of the order $q_{max}$ to attache  along
its $q$-face is given by
\begin{equation}
P(q_{max},q;t)= \frac{c_q(t)e^{-\nu (q_{max}-q)}}{\sum _{q=0}^{q_{max}-1}c_q(t)e^{-\nu (q_{max}-q)}} \ .
\label{eq-pattach}
\end{equation}
where $\nu$ is the chemical affinity parameter, see
Ref. \cite{we-SciRep2018} for a detailed description.
The $c_q(t)$ is the number of geometrically compatible locations for
docking a simplex of the order $q$,  
counted on the entire growing structure at the time $t$. 
\begin{figure}[htb]
\centering
\begin{tabular}{cc} 
\resizebox{18pc}{!}{\includegraphics{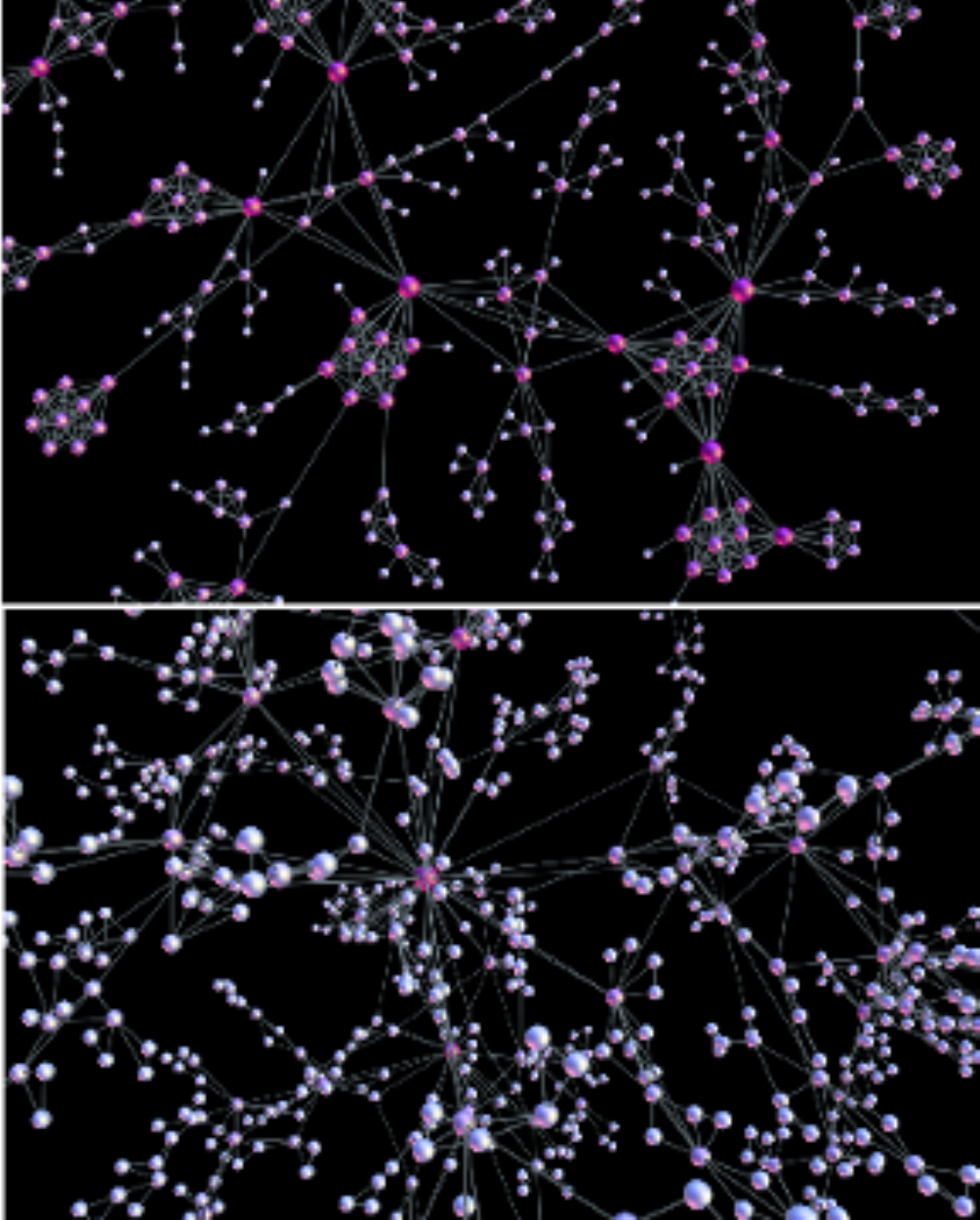}}\\
\end{tabular}
\caption{Top: Zoom-in nanoparticle network self-assembled by preformatted groups of particles as cliques of different sizes $n\in[2,10]$ for a negative chemical affinity $\nu$ (significant repulsion between the cliques). Bottom: Self-assembled network of triangles fo $\nu=0$ (geometrically constraint self-assembly).}
\label{fig:nets}
\end{figure}
The geometrical factor is weighted by the chemical affinity of the system towards the addition of  $n_a=q_{max}-q$ new nodes. Specifically, for a large $\nu<0$, the system favours the addition of nodes. Thus the cliques preferably share a single node (see an example in Fig.\ \ref{fig:nets} top panel). In the opposite limit, with increasing  $\nu >0$, the cliques progressively share their larger face; eventually, a single node can be added, leading to highly compact structures, as shown in   \cite{we-SciRep2018,we-PRE2020}.
In the case when $\nu=0$,  the process is governed by the geometric
compatibility factors alone.  Consequently, each face can be
shared with a finite probability, depending on its size and the actual
network structure. See online demo on this link \cite{we-ClNets-applet}.

\begin{figure}[!htb]
\centering
\begin{tabular}{cc} 
\resizebox{18pc}{!}{\includegraphics{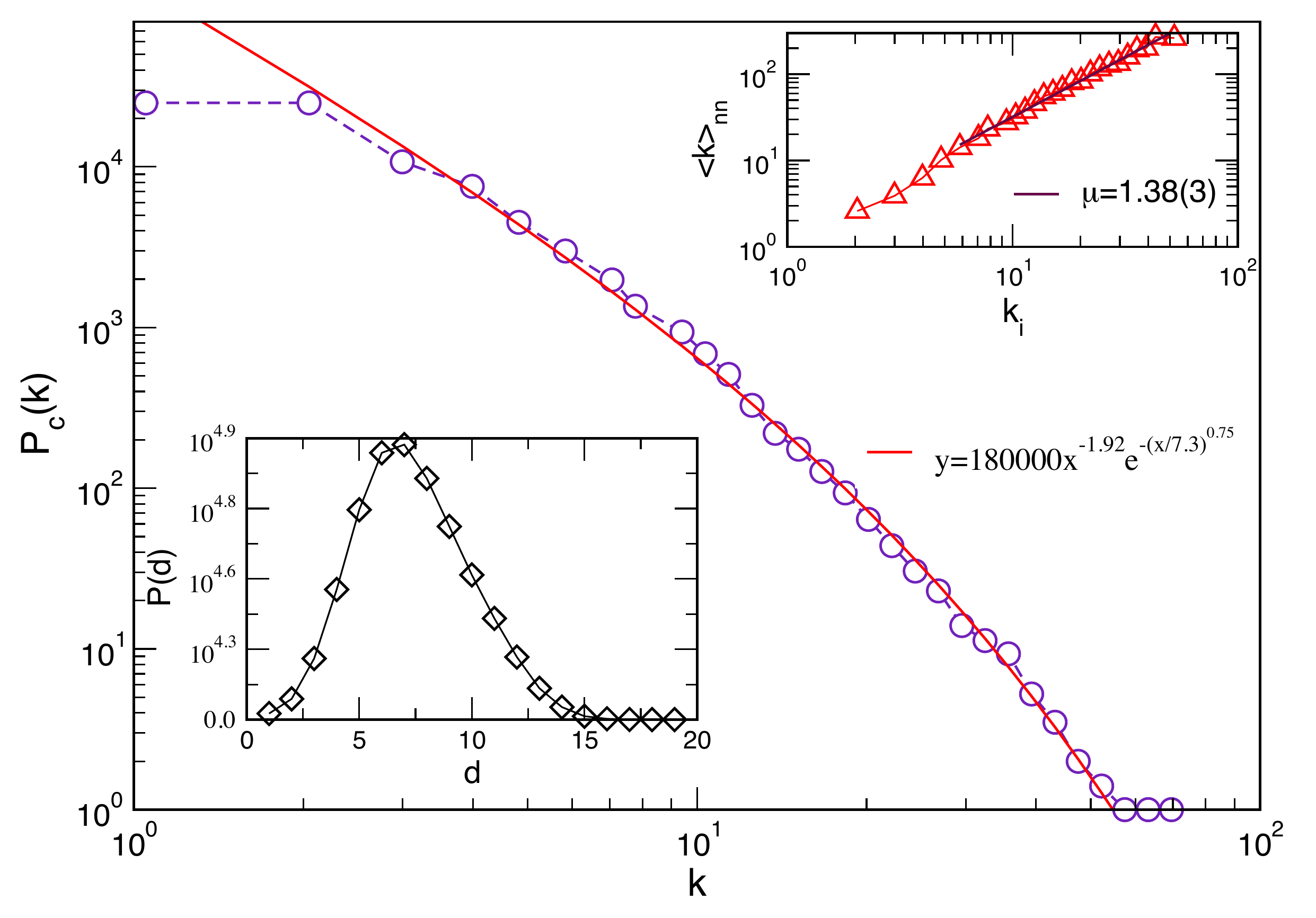}}\\
\end{tabular}
\caption{For the network of self-assembled triangles: Sample averaged
  cumulative distribution $P_c(k)$ of the degree $k$  (central panel)
  and the assortativity: the average degree of neighbour nodes plotted
  against the node's $i$ degree (top inset), samples with 5000 nodes. Bottom inset: The distribution $P(d)$ of the shortest-path distances $d$, for the assembly with 1000 nodes. }
\label{fig:triangdistr}
\end{figure}

For this work, we grow an assembly of triangles under
the strictly geometrical compatibility conditions ($q_{max}=n-1=2$, and
$\nu=0$).  A close-up of the grown structure is depicted in the bottom panel of Fig.\ \ref{fig:nets}; it possesses 1000 nodes, 1737 edges, and 738 triangles.
As shown in \cite{we-SciRep2018,we-PRE2020}, these
networks exhibit a broad distribution of the node's degree,  
assortative mixing, and a wide range of shortest-path distances
between nodes (see Fig.\ \ref{fig:triangdistr}). They also have the hyperbolic geometry (a negative curvature in the shortest-path metric space), precisely, they are 1-hyperbolic.  The origin of the small hyperbolicity parameter in these networks is in  the attachment
of new clique via shared face with an existing clique.
Thus, the distance between individual cliques, which are ideal hyperbolic structures ($\delta_{clique}=0$),   is minimal leading  to  $\delta_{clique}+1$-hyperbolic clique-complexes
\cite{HB-BermudoHBviasmallerGraph2013,Hyperbolicity_cliqueDecomposition2017}. We note that this attachment rule also prevents the appearance of
holes in networks and that the order of a simplicial complex is equal
to the order of the largest building simplex. 

To study the magnetisation-reversal dynamics, here we adopt the
network as shown in Fig.\ \ref{fig:nets} bottom panel,  and
attach an Ising spin to each node (nanoparticle). 
The interactions among pairs of spins are enabled by the network
links (adjacency matrix). Whereas, the triangle-based interactions
are defined as \textit{tri-spin interactions among spins situated on a given triangle}. In general, the triangle-adjacency matrix can be
inferred by identifying all triangles, i.e., by performing the  
Q-analysis of the network
\cite{Qanalysis1,we-SciRep2018,we-PRE2020}. In the present
case, by growing the network in time steps $t$, we keep track of
the nodes $i_1,i_2,i_3$ belonging to the added triangle.  For example,
in the first 11 steps, we can observe the following sequence of events
(the variables in curly brackets are \{$t$,$N_t$,$\Sigma_t$,$n$,
$n_a$, $i_1,i_2,i_3$\}, where $n$ and $n_a$ are defined
above, and $N_t$ and  $\Sigma_t$ stand for 
the number of nodes, and simplexes and faces, respectively.): 
\begin{verbatim}
{1 3 7 3 2 1 2 3},
{2 4 11 3 1 1 3 4},{3 5 15 3 1 2 3 5},
{4 6 19 3 1 1 3 6},{5 7 23 3 1 3 4 7},
{6 9 29 3 2 4 8 9},{7 10 33 3 1 3 7 10},
{8 11 37 3 1 8 9 11},{9 12 41 3 1 3 10 12},
{10 14 47 3 2 7 13 14},{11 16 53 3 2 12 15 16}
\end{verbatim}

\section{Spin reversal dynamics under triangle-based interactions\label{sec:results}}
As mentioned in the Introduction, the antiferromagnetic spin-spin interactions in conjunction with a complex geometry of the assemblies provide ideal conditions for the spin frustration effects \cite{HOC_BHN_weEntropy2020}. 
Here, we study the spin reversal dynamics driven by the
external field $h_{ext}$ on an assembly of triangles, cf.  Fig.\
\ref{fig:nets}. Each pair of adjacent spins obeys antiferromagnetic
interaction, whereas the higher-order interaction involves the spins based on the same triangle. The Hamiltonian is given by
\begin{equation}
{\cal{H}} = (\alpha -1)\sum _{i,j}J_{ij}S_iS_j
-\alpha\sum_{<i,j,k>}K_{ijk}S_iS_jS_k- h_{ext}\sum _iS_i 
\label{eq:ham}
\end{equation}
where the Ising spins $S_i=\pm 1$ are associated with the network
vertices; the summation in the first term is over all pairs of spins that have a link in the network's adjacency matrix, whereas the sum in the second term runs over the identified triangles, as described above.
The spin-spin interaction  $J_{ij}=-J$ is antiferromagnetic, whereas
$K_{ijk}=K_3$ for the simplex-based interaction, where
we consider both $K_3>0$ as well as $K_3<0$ cases. The parameter $\alpha\in[0,1]$ is varied to provide a relative balance between the pairwise and higher-order interactions. We use the
dimensionless units and set $J=1$ and $K_3=\pm1$. The magnetisation
(in Bohr magnetons $\mu_B$) is determined by the balance of the up and
down oriented spins, $M=(N_+ -N_-)/N \in [-1,1]$. Meanwhile, the
external field $h_{ext}$  (in the units $J\mu_B$) varies in the range $[-h_{max},+h_{max}]$.
The range relevant for the process is related to the maximum number of neighbours of the vertices.

\begin{figure}[!htb]
\centering
\begin{tabular}{cc} 
\resizebox{18.8pc}{!}{\includegraphics{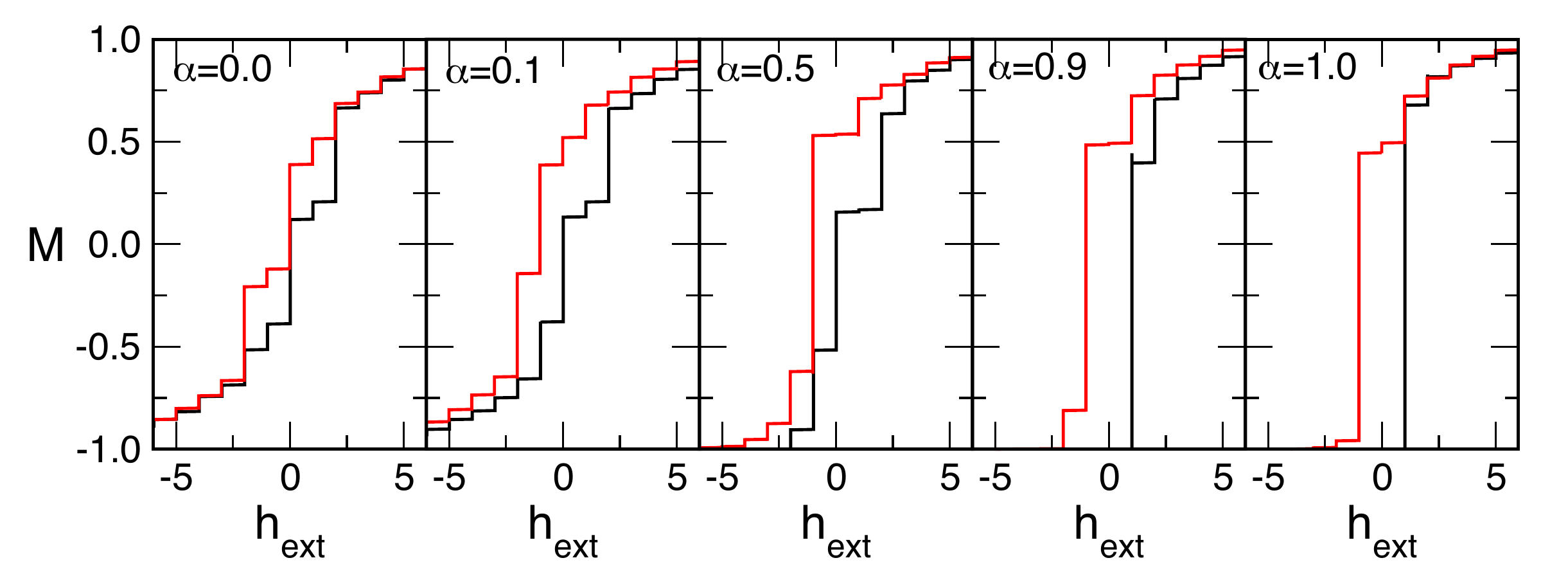}}\\
\end{tabular}
\caption{Hysteresis loop in field-driven magnetisation reversal on the self-assembled network of triangles, see Fig.\ \ref{fig:nets}; different panels are for varied $\alpha$ from 0, corresponding to
  pure pairwise interaction to 1, entirely triangle-based
  interactions, see Eq.\ (\ref{eq:ham}). Different colour indicates ascending and descending branches. The case with $K_3>0$ is shown.}
\label{fig:hloops}
\end{figure}

The process is driven by slow ramping of the external field along an ascending/descending branch of the hysteresis loop. A zero-temperature dynamics is applied, as it is a widely accepted approach in the study of Barkhausen noise in disordered magnetic systems, see \cite{we_BHN_SciRep2019} and references therein. A small increase of the field can trigger an avalanche of spin reversal, during which each spin aligns along its local fields to minimise the energy.  In the disordered ferromagnets, an avalanche of spin flips stops when the magnetic defects pin the domain wall of the expanding domain of reversed spins. In contrast, in the spin assembles that we are studying here, there are no magnetic defects. Instead, the geometry of the network alone governs the progression of the spin reversal process
(see also \cite{we_BHN_sfnets,HOC_BHN_weEntropy2020}).
Specifically, for the ascending branch, we start from a state with all spins down, and a large negative filed $h_{ext}=-h_{max}$ is applied. $h_{max}\leq k_{max}$ correlates with the number of connections of the leading hub in the network. The field changes are  \textit{adiabatic}, which means that the field is kept fixed during the avalanche propagation.  The field is increased in small steps $\Delta$ until it reaches the other limit $h_{ext}=+h_{max}$.  Then the field is ramped by slow decreasing until it reaches $-h_{max}$ again, to complete the loop. Apart
from the current value of the external field, the local field at the
vertex $i$ is given by the contribution of the actual states of the
neighbouring spins according to the interactions in (\ref{eq:ham}).
In the present case, due to the fixed strength of the interactions, the
local filed changes in the integer values, which depend on the node's
pairwise and triangle-based connectivity,  modified by the parameter $\alpha$. In analogy to modelling the charged domain-wall motion during the ferroelectric switching \cite{BHN_FEnatt2016},  the frustration effects within the zero-temperature dynamics are emulated by the probabilistic alignment between the spin and its local field with a probability $c<1$. 
The results for the hysteresis loop are shown in Fig.\ \ref{fig:hloops} for several values of the parameter $\alpha$, interpolating between the strictly pairwise antiferromagnetic interactions, for $\alpha=0$, and strictly triangle-based interactions, for $\alpha=1$. 

\begin{figure}[!htb]
\centering
\begin{tabular}{cc} 
\resizebox{18pc}{!}{\includegraphics{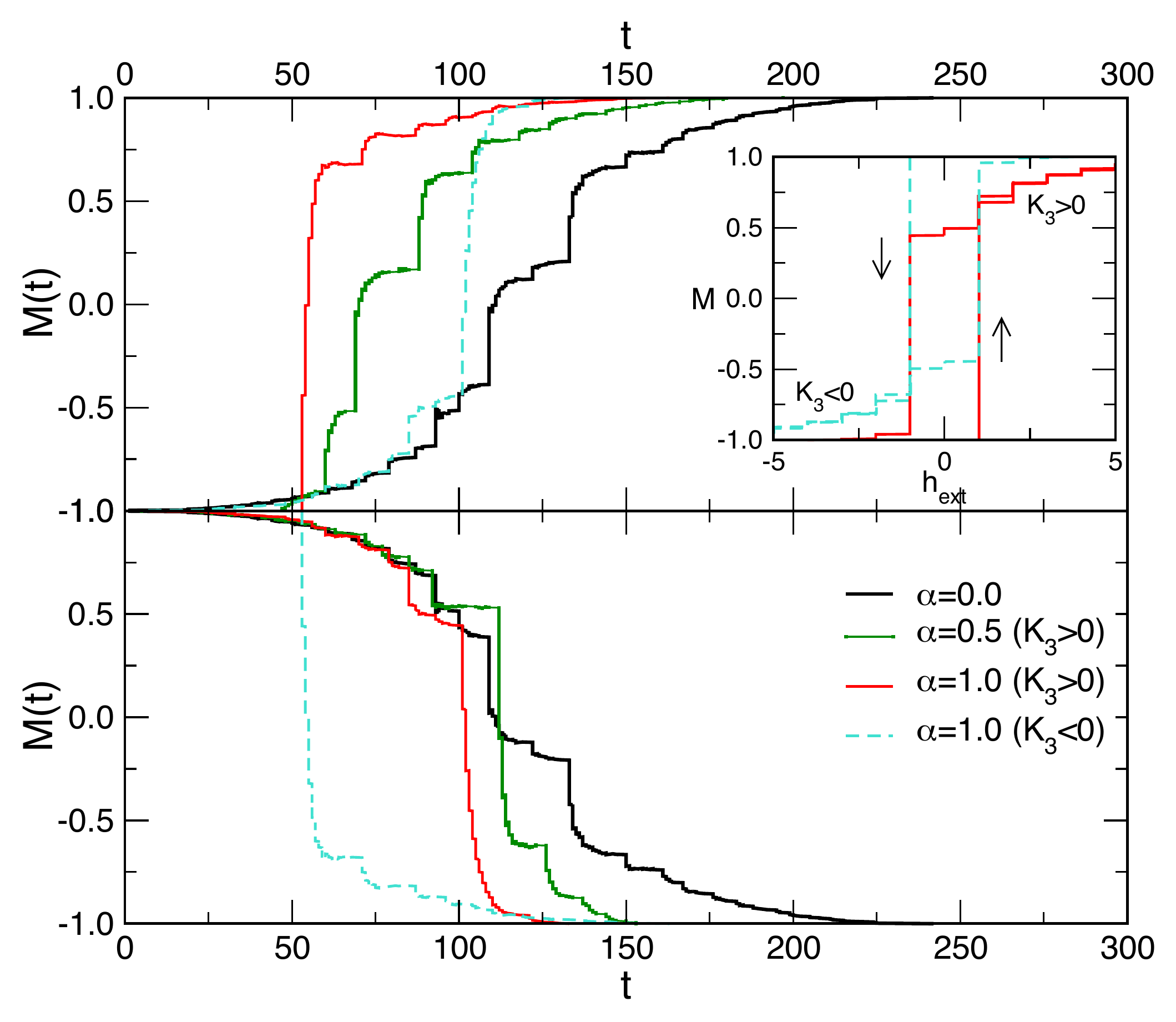}}\\
\end{tabular}
\caption{Magnetisation versus time in the field-driven reversal processes on the
  self-assembled network of triangles, see Fig.\ \ref{fig:nets}; different
  panels are for varied $\alpha$ from 0, corresponding to
  pure pairwise interaction to 1, entirely triangle-based
  interactions, see Eq.\ (\ref{eq:ham}). Ascending and descending
  branches are shown of separate panels for a better view.
  Inset: Hysteresis loop  for $\alpha =1$ (only triangle-based interactions) with $K_3>0$ and $K_3<0$. }
\label{fig:Mt}
\end{figure}

As Fig.\ \ref{fig:hloops} shows, in the absence of the
higher-order interactions, the hysteresis loop si symmetrical and,
typically for the antiferromagnetic samples,  splits into the positive and negative part. Moreover, the narrow tails at both high positive and high negative fields indicate the effects of a robust topological disorder, which permits only small avalanches of the reversed spins. The appearance of the higher-order interactions, however, induces a broadening and breaks the symmetry of the loop.  In approaching the limiting case for $\alpha=1$, the triangle-based interactions cause an entirely different shape of the hysteresis loop with a large rectangular part (resembling a ferromagnetic case). However, it also shows a disorder-conditioned tail, but on one side of the loop only. The implicated side depends on the sign of the $K_3$, see also the inset in Fig.\ \ref{fig:Mt}.
Consequently, the increase of the magnetisation in time differs for
the ascending and descending branch of the hysteresis whenever $\alpha$ is finite, as shown in the main panel of Fig.\ \ref{fig:Mt}. For $\alpha =1$, a  similarity occurs between the magnetisation curves in different branches,  but for the reversed sign of $K_3$. More precisely, the ascending line for $K_3>0$ is a mirror reflection of a descending line for $K_3<0$, and also the ascending line for the case $K_3<0$
mirrors the descending line for $K_3>0$, cf. full red lines and the
corresponding dashed pale lines in Fig.\ \ref{fig:Mt}.

\begin{figure}[!htb]
\centering
\begin{tabular}{cc} 
\resizebox{18pc}{!}{\includegraphics{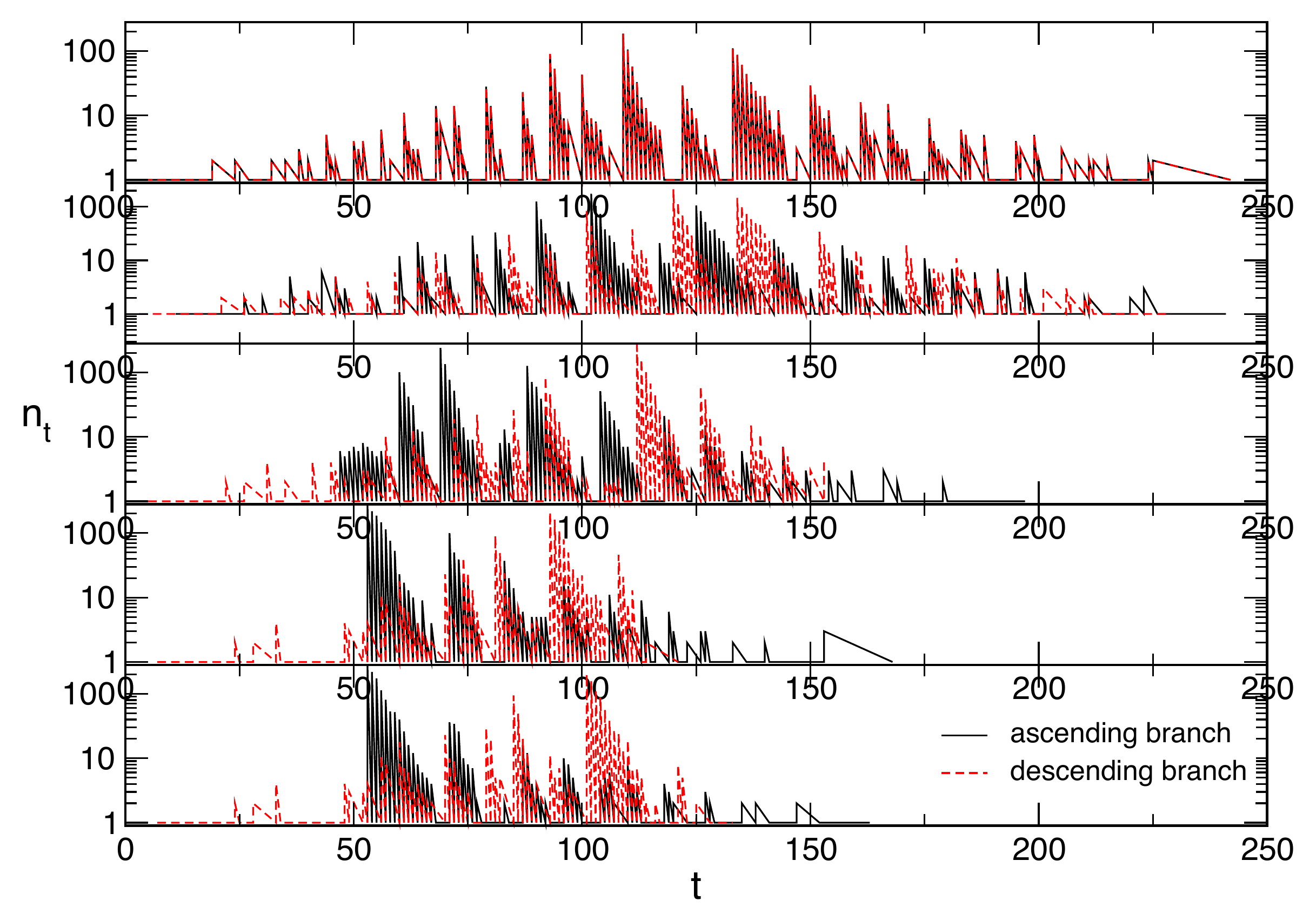}}\\
\resizebox{18pc}{!}{\includegraphics{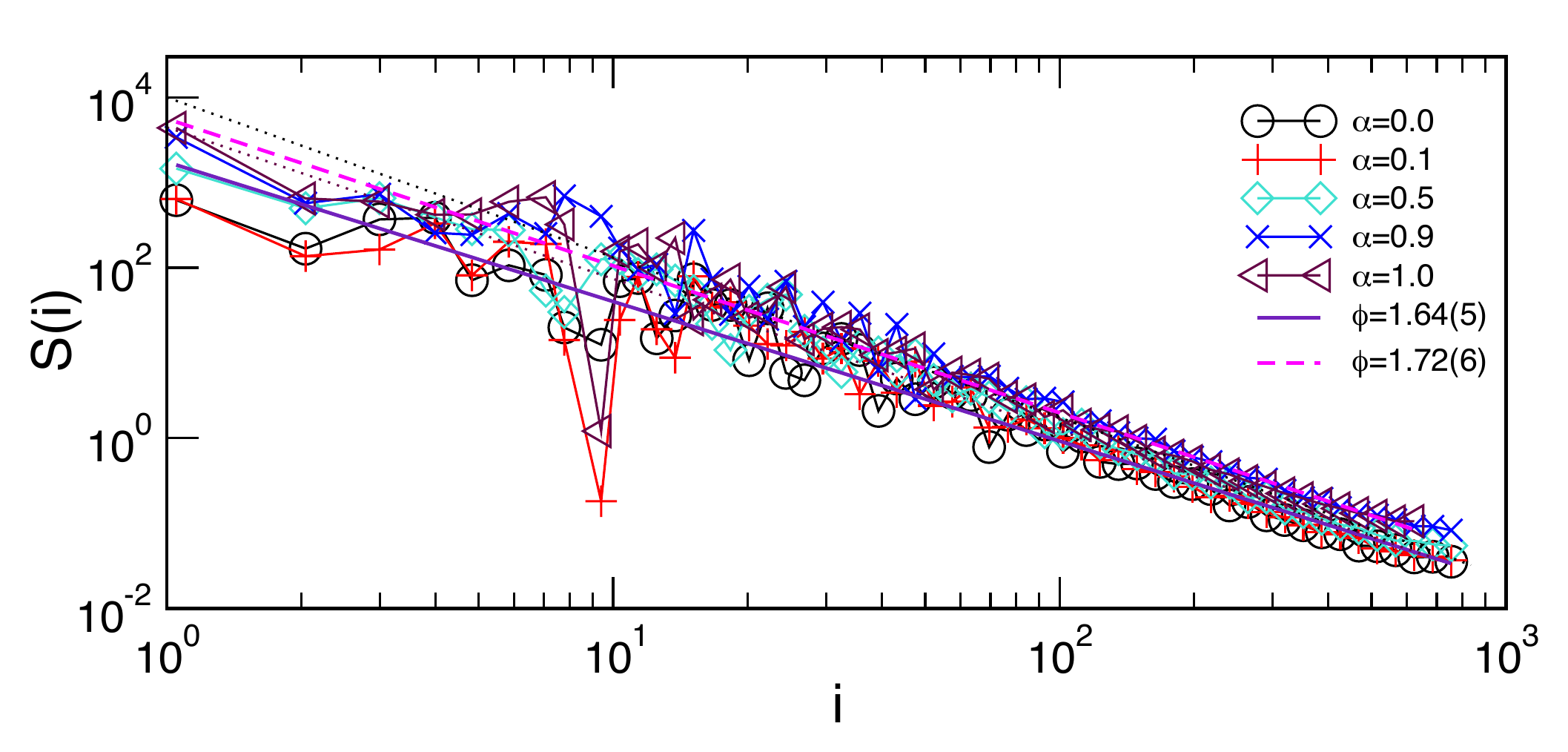}}\\
\end{tabular}
\caption{Barkhausen noise signal $n_t$ against time $t$ in the field-driven reversal  processes on the self-assembled network of triangles, see Fig.\ \ref{fig:nets};  five panels from the top  are for different  $\alpha \in[0,1]$, from the pure pairwise interaction  to entirely triangle-based
  interactions as in Fig.\ \ref{fig:hloops}. Differences in the ascending and descending branches signal are indicated by colour.  Bottom panel: Power spectrum of the signals (ascending branch) and their power-law fits, the
maximum and minimum slopes are displayed.}
\label{fig:noise}
\end{figure}

\begin{figure}[!htb]
\centering
\begin{tabular}{cc} 
\resizebox{18pc}{!}{\includegraphics{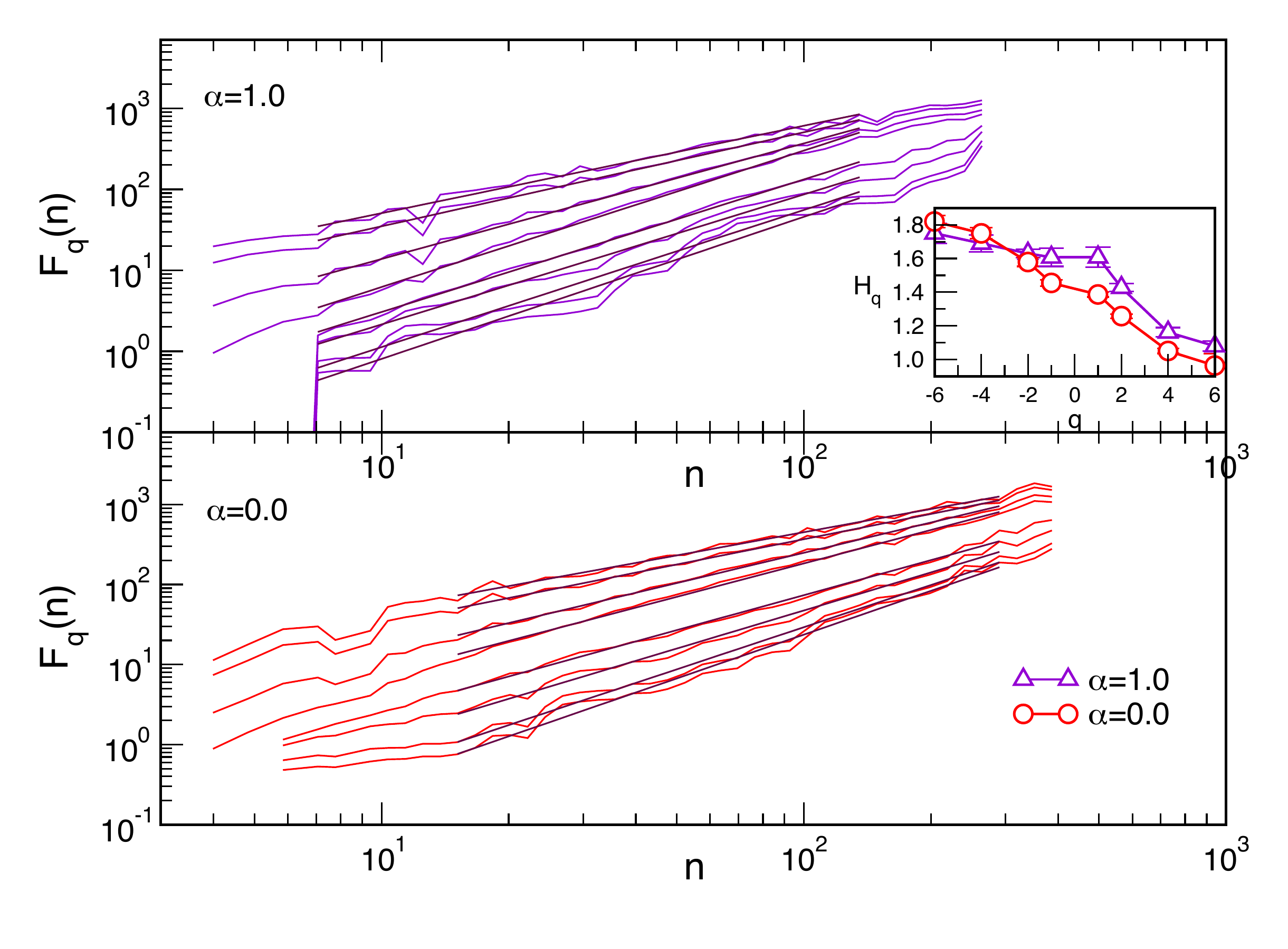}}\\
\end{tabular}
\caption{Fluctuation function $F_q(n)$ vs. the segment length $n$ for
  $q=$ 6,4,2,1,-1,-2,-4,-6 (top to bottom curve) for the ascending branch
  in the limiting cases $\alpha=1.0$ and $\alpha =0.0$. Inset: 
  values of the generalised Hurst exponent $H_q$ corresponding to the
  slope of the straight line indicated on each curve.  }
\label{fig:Fq}
\end{figure}

The magnetisation fluctuations comprising the Barkhausen noise signals during the reversal processes for different $\alpha$ are shown in Fig.\ \ref{fig:noise}.  The bursting events (avalanches) triggered by the changed of the external field have a characteristic triangular shape caused by the underlying geometry in the absence of
disorder. With the increased $\alpha$, we notice that the signal
becomes shorter as well as its most massive avalanches occur much earlier, following the shape of the hysteresis loop in Fig.\ \ref{fig:hloops}. However, they all exhibit self-similarity with the long-range temporal correlations, as captured by the corresponding power-law behaviour of the power spectrum, $S(i)\sim i^{-\phi}$, cf.\ lower panel in Fig.\ \ref{fig:noise}. 
Furthermore, we find that these noise signals possess the multifractal
features, similar to the Barkhausen noise in disordered ferromagnets 
\cite{we_BHN_MFR2016}. More precisely, the fluctuation function
$F_q(n)$ dependence on the segment length $n$ of the signal exhibits a
power-law behaviour with a spectrum of the exponents. In particular, dividing the signal in $N_s$ segments and having defined 
 the standard deviation $F(\mu,n) =\left\{
  \frac{1}{n}\sum_{i=1}^n[Y((\mu-1)n+i)-y_\mu(i)]^2\right \}^{1/2}$
around a local trend $y_\mu(i)$ in each segment $\mu=1,2\cdots N_s$,
we have 
\begin{equation}
F_q(n)=\left\{\frac{1}{2N_s}\sum_{\mu=1}^{2N_s} \left[F^2(\mu,n)\right]^{q/2}\right\}^{1/q} \sim n^{H(q)} 
\label{eq:Fq}
\end{equation}
Here, a generalised Hurst exponent $H_q$ depends on the amplification factor $q$, as shown in Fig.\ \ref{fig:Fq}. Intuitively, this means that the small fluctuations (amplified when $q<0$) along the signal have different scaling properties than the large fluctuations ($q>0$). In contrast, for the
monofractal signals, $H_q=H_2$ for all $q$, where $H_2$ is the
standard Hurst exponent. We note that a small variation in $H_2$ agree
with small variations in the exponent of the power spectrum, hence
 the expected scaling relation $\phi =2H_2-1$ is roughly satisfied (within
the error bars).

 The critical behaviour at the hysteresis loop in disordered ferromagnets is often related to the occurrence of a critical disorder point or a critical line \cite{we_BHN_SciRep2019} in the parameters space. Even though signatures of the self-organised criticality are present. For a recent review of systems exhibiting self-organised criticality in
 \cite{SelfOrganizedCriticalitySystems2013} and references therein. 
As stated above,  the avalanching process that we study here is governed by the geometry of the assembly without any magnetic disorder. 
Nevertheless, Fig.\ \ref{fig:noise} and Fig.\ \ref{fig:Fq} demonstrate robust features of criticality in the magnetisation fluctuations.  A detailed analysis of scale invariance of the avalanches remains for separate work. 

\section{Conclusions} We have briefly presented the subjects of current compelling research directions aiming to understand the role of the higher-order connectivity and interactions in the architecture and functional properties of complex systems. As a representative example, we have studied the nano-network grown by self-assembled simplexes (triangles), and the field-driven magnetisation reversal on them. With the antiferromagnetic interactions among the adjacent spin pairs and the triangle-based three-spin interactions, the underlying architecture enables the geometric frustration effects that have a pronounced impact on the hysteresis loop in these assemblies. By balancing between the pairwise and triangle-based interactions, the geometric frustration effects are gradually overridden by structural inhomogeneities due to the architecture of simplicial complexes. Moreover,  the observed fluctuations of the magnetisation show a robust signature of self-organised criticality.  The avalanching processes are conditioned solely by the geometry of the assembly without any magnetic disorder. These findings are in the line of research of complex nano-assembled materials with emerging functional properties.

\acknowledgments
Work supported by the Sovenial Research Agency under the Program
 P1-0044.

\end{document}